\documentclass[10pt,aps,twocolumn]{revtex4-1}
\usepackage{graphicx}
\usepackage{amssymb}
\usepackage{amsmath}
\usepackage{appendix}
\usepackage{comment}

\def\p{\partial}
\def\i{\imath}
\def\j{\jmath}

\def\br{\mathbf{r}}
\def\bs{\mathbf{s}}
\def\bj{\mathbf{j}}
\def\bk{\mathbf{k}}
\def\bn{\mathbf{n}}

\def\bp{\mathbf{p}}

\def\bv{\mathbf{v}}
\def\bz{\mathbf{z}}
\def\bx{\mathbf{x}}
\def\bA{\mathbf{A}}

\def\BN{\boldsymbol{\nabla}}
\def\BE{\boldsymbol{\varepsilon}}

\def\BT{\boldsymbol{\tau}}
\def\mA{\mathcal{A}}

\def\mS{\mathcal{S}}
\newcommand{\rf}[1]{(\ref{#1})}
\newcommand{\al}[1]{\begin{aligned}#1\end{aligned}}

\newcommand{\eq}[1]{\begin{equation}#1\end{equation}}

\usepackage{mdframed}

\begin{document}

\title{Quantum Hydrodynamics of Vorticity}

\author{Yaroslav Tserkovnyak}
\author{Ji Zou}
\affiliation{Department of Physics and Astronomy, University of California, Los Angeles, California 90095, USA}

\begin{abstract}
We formulate a quantum theory of vorticity (hydro)dynamics on a general two-dimensional bosonic lattice. In the classical limit of a bosonic condensate, it reduces to conserved plasma-like vortex-antivortex dynamics. The nonlocal topological character of the vorticity flows is reflected in the bulk-edge correspondence dictated by the Stokes theorem. This is exploited to establish physical boundary conditions that realize, in the coarse-grained thermodynamic limit, an effective chemical-potential bias of vorticity. A Kubo formula is derived for the vorticity conductivity|which could be measured in a suggested practical device|in terms of quantum vorticity-flux correlators of the original lattice model. As an illustrative example, we discuss the superfluidity of vorticity, exploiting the particle-vortex duality at a bosonic superfluid-insulator transition.
\end{abstract}

\maketitle

\section{Introduction}

It is now well appreciated that electrically insulating materials may exhibit a wealth of neutral transport phenomena. The underlying conserved quantities may emerge out of certain symmetries associated with the microscopic spin \cite{zuticRMP04,*maekawaBOOK17} or pseudospin (e.g., valley \cite{rycerzNATP07,*castroRMP09}) degrees of freedom or, alternatively, topology of the collective dynamics \cite{tserkovJAP18}. The former scenario, which has been most thoroughly exploited in the field of spintronics \cite{zuticRMP04}, concerns the spin angular momentum along a high-symmetry axis of the pertinent heterostructure. As the full axial symmetry is inevitably broken in spin space, at some level, the resultant spin hydrodynamics is always approximate, being useful only on some finite time and/or length scales. The topological hydrodynamics, on the other hand, is potentially more robust, as it is rooted in the topological structure of the dynamical variables rather than any specific structural symmetries \cite{tserkovJAP18}.

At the heart of this are conservation laws constructed out of a topological invariant of the dynamical field configurations, such as the winding number of a one-dimensional XY model \cite{takeiPRL14,*kimPRB15br} or a superfluid \cite{halperinIJMPB10} or the skyrmion number \cite{belavinJETPL75} of a two-dimensional Heisenberg model \cite{ochoaPRB16}. These topological invariants are endowed by the homotopic properties of the smooth field configurations of the bulk, following, for example, the $\pi_n(S^d)=\mathbb{Z}$ group-theoretic structure of the $n$th homotopy on $d$-sphere, when $n=d$ \cite{nakaharaBOOK03}. The integer on the right-hand side here counts the conserved topological ``charge" that can be associated with the dynamical fields. Being conserved in the bulk, this effective charge can, nonetheless, flow in and out of the medium through its boundaries, which hints at a possibility of its control: The (nonequilibrium) boundary conditions could be devised to bias injection of the topological charge of certain sign and, reciprocally, detect its outflow elsewhere \cite{tserkovJAP18}.

This points to a conceptual possibility of assigning a bulk conductivity to the topological charge fluctuations in the material, which could potentially extend many of the useful and intuitive notions associated with the charge conductivity to broad classes of insulating materials. Both device possibilities and novel transport probes of fundamental material properties could then be expected to arise hand in hand. The outlook may, however, be hindered by one key approximation underlying such hydrodynamic constructions: The overarching topological invariant is a property of a low-energy sector of the theory, with pathological excursions between different topological sectors possible in principle. In the case of the winding dynamics, such excursions are known as phase slips, which are central to understanding low-dimensional superfluidity and superconductivity \cite{halperinIJMPB10}. Skyrmions, likewise, can be created and annihilated by local fluctuations \cite{diazCM16,*derrasPRB18}. Such detrimental phase-slip-like events, which ultimately relax any topological configuration towards the global equilibrium, can originate either at the atomistic level, where the coarse-grained treatment of the smooth field theory breaks down, or more macroscopically, where the dynamical variables deviate significantly from their presumed low-energy manifold. Even when such processes are rare, in the limit when they are exponentially suppressed by a large energy barrier, their existence poses a technical challenge in formulating a transport theory. After all, there is no strict continuity equation for the topological charge at the microscopic quantum level, unless we formally separate and eliminate the phase-slip events. Depending on the exact model, parameters, and ambient temperature, furthermore, there may be a plethora of scenarios for the phase-slip dynamics \cite{halperinIJMPB10}, which could diminish the utility of the topological conservation law.

In this paper, we formulate a topological hydrodynamics that is based on a robust continuity equation immune to all these issues. Its formal distinguishing feature is in the conserved quantity that is related to the field homotopy defined on the boundary rather than the bulk, which, nevertheless, determines a conserved bulk quantity according to a Stokes theorem \cite{zouPRB19}. We demonstrate this general idea by the ground-up construction of a quantum vorticity hydrodynamics in $2+1$ dimensions, offering links to quantum spintronics \cite{tserkovJAP18}, particle-vortex dualities in many-body systems \cite{wenBOOK04,*seibergAP16,*senthilCM18}, and quantum turbulence \cite{paolettiARCMP11,*madeiraCM19}.

The paper is structured as follows: In Sec.~\ref{cvd}, we briefly review a vorticity conservation law emerging in two-dimensional superfluids, which reduces to the simple vortex-antivortex counting in the appropriate strongly-ordered limit. In Sec.~\ref{qvd}, a corresponding quantum theory is constructed on a generic bosonic lattice, which mimics the aspects of the classical vorticity dynamics. The conservation law is now formulated, for an arbitrary Hamiltonian, at the level of the microscopic Heisenberg equation of motion. In this section, we also discuss boundary conditions for injecting and detecting the vorticity flows, formulate a field-theoretic Kubo formula for calculating the associated conductivity, and apply it to some illustrative examples. A summary and outlook are offered in Sec.~\ref{so}.

\section{Classical vorticity dynamics}
\label{cvd}

A conventional superfluid condensate can be described by a complex-valued order parameter $\phi$. The corresponding scalar field $\phi=\sqrt{n}e^{i\varphi}\in\mathbb{C}$ (where $n\geq0$ is the condensate density and $\varphi\in\mathbb{R}$ is its phase) residing in $2+1$ dimensions, $\phi(\br,t)$, realizes an $\mathbb{R}^2\to\mathbb{C}$ mapping, at any given time $t$. These field textures are devoid of point defects, as the fundamental homotopy group of the complex plane is trivial, $\pi_1(\mathbb{C})=1$. Such two-dimensional textures are, furthermore, all topologically equivalent, having fixed the boundary profile of $\phi$ on a simply-connected patch of $\mathbb{R}^2$, which is reflected in the fact that $\pi_2(\mathbb{C})=1$. Despite this, a smooth vector field defines a \textit{topological hydrodynamics} \cite{tserkovJAP18} governed by the continuity equation $\p_\mu j^\mu=0$ (with the Einstein summation implied over the Greek letters: $\mu=0,1,2\leftrightarrow t,x,y$), where
\eq{
j^\mu=\frac{\epsilon^{\mu\nu\xi}\p_\nu\phi^*\p_\xi\phi}{2\pi i}\,.
\label{jmu}}
Here, $\epsilon^{\mu\nu\xi}$ is the Levi-Civita symbol. For a rigid texture sliding at a velocity $\bv$, $\bj=\rho\bv$, where $\rho\equiv j^0$ and $\bj=(j^x,j^y)$. For a sharp vortex in an ordered medium with the free energy minimized by a finite $n$, $\rho\approx n\delta(\br-\br_0)$, where $\br_0$ is the vortex-\textit{core} position where $n$ vanishes. Fixing a finite magnitude of the scalar field, the homotopy group would in this case become $\pi_1(S^1)=\mathbb{Z}$, counting essentially the number of vortices in the system.

The conserved quantity can be recast as a fictitious flux ($\bz$ is the z-axis unit vector):
\eq{
\rho=\frac{\bz\cdot\BN\phi^*\times\BN\phi}{2\pi i}=\frac{\bz\cdot\BN\times\bA}{2\pi}\,,
\label{rho}}
associated with the \textit{gauge field}
\eq{
\bA=-i\phi^*\BN\phi\,.
\label{A}}
Applying Green's theorem, we then see that the conserved \textit{topological charge} within a patch $\mS$,
\eq{
Q\equiv\int_\mS d^2r\,\rho=\oint_{\p\mS}\frac{d\br\cdot\bA}{2\pi}=\oint_{\p\mS}\frac{d\varphi}{2\pi}\,n\,,
\label{Q}}
is associated with the phase winding around its boundary $\p\mS$. This reveals the geometrical meaning of the conservation law: The charge $Q$ in the bulk can change only in response to a vorticity flow through the boundary.

Alternative to Eq.~\eqref{jmu}, it might be tempting to write the current density associated with field dynamics as
\eq{
\bj=\frac{\bz\times\p_t\bA}{2\pi}\,,
\label{j}}
from which $\p_t\rho+\BN\cdot\bj=0$ immediately follows. This current, however, differs from the more physical definition \rf{jmu} by a nonlocal (divergenceless) shift, which would spoil our energetic and Kubo considerations below.

Note that a similar conservation law, with the current \eqref{j}, applies to any other density $\rho$ that can be written as Eq.~\eqref{rho} in terms of some field $\bA(\phi)$. Our choice \eqref{A} for this field merely results in the physical interpretation of the conservation law in terms of the vorticity \eqref{Q} dynamics. This is particularly relevant for ordered condensates, where the vortices become quantized in terms of the elementary charges $Q=\pm1$, interacting via a two-dimensional (long-range) electrostatic coupling \cite{kosterlitzJPC73}.

Let us  consider some other simple examples of conserved hydrodynamics associated with different choices for the field $\bA(\phi)$, which defines the topological charge \rf{rho}. First, we note that there is a \textit{gauge} freedom in defining $\bA(\phi)$: $\bA(\phi)\to\bA(\phi)+\BN f(\phi)$, in terms of an arbitrary function $f(\phi)$, which leaves $\rho$ unchanged. This is why the first-order (derivative) fields like $\bA=\BN n$ are physically inconsequential. Indeed, the corresponding conserved quantity $Q=\oint dn=0$, for a smooth field $\phi(\br)$. Perhaps the simplest nontrivial example is given by $\bA=2\pi({\rm Re}\,\phi,{\rm Im}\,\phi)$, which results in $Q=\oint d\br\cdot\bA/2\pi$. The largest possible charge $Q$ within a given simply-connected region (of radius $\sim R$), for a fixed $n$, corresponds to placing a single vortex in the interior, which gives $Q/\sqrt{n}\to 2\pi R$, the circumference of the region. The corresponding density $\rho/\sqrt{n}\sim2/R$ vanishes in the thermodynamic limit of $R\to\infty$. We, therefore, conclude that the vorticity density generated by the gauge field \rf{A} gives the simplest topological charge that can result in a physically meaningful \textit{extensive} hydrodynamics.

\section{Quantum vorticity dynamics}
\label{qvd}

\begin{figure}[!th]
\includegraphics[width=0.7\linewidth]{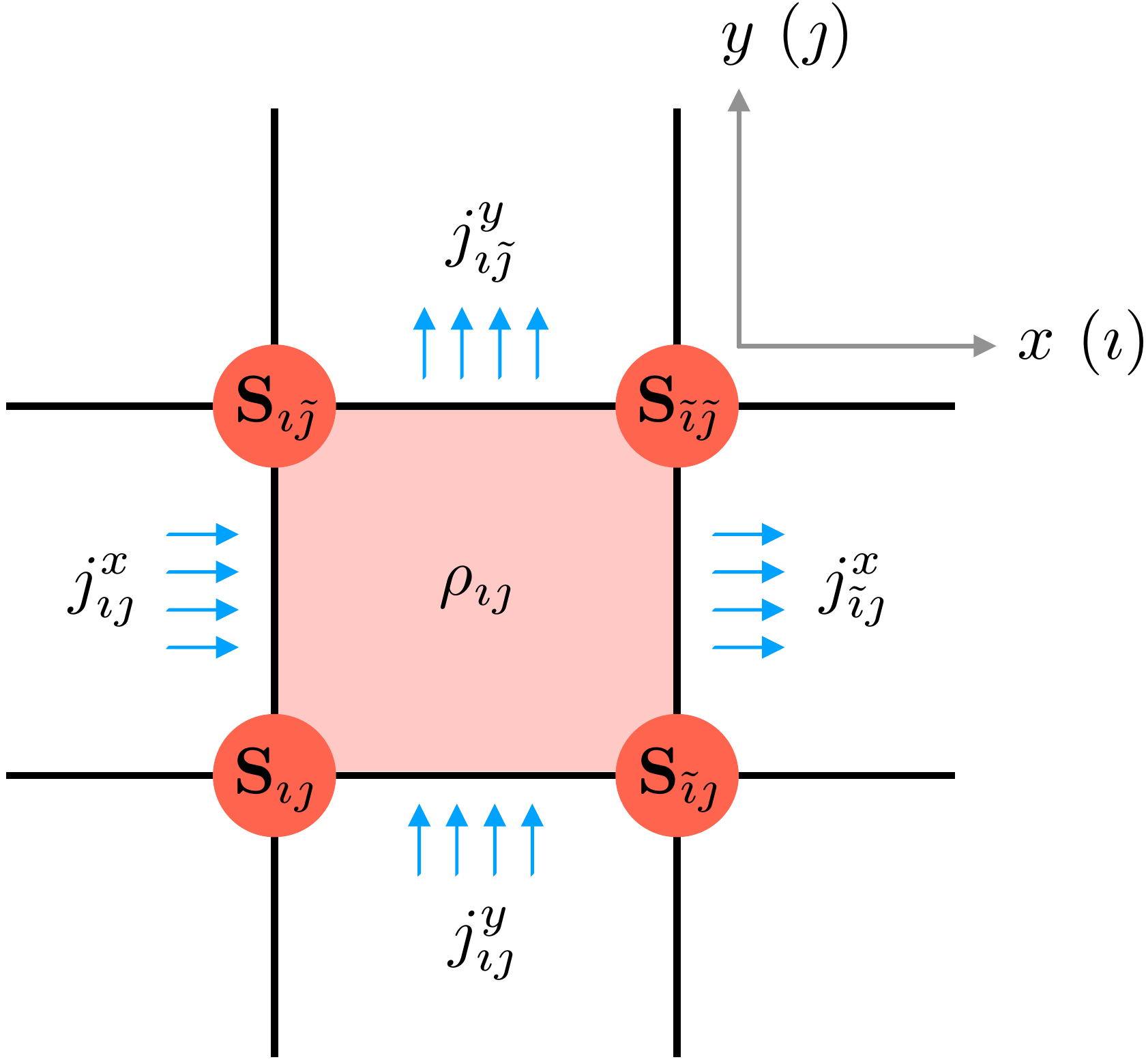}
\caption{A quantum bosonic lattice described by an arbitrary Hamiltonian $H$. $\Phi_{\i\j}$ is the bosonic field operator at site $\i\j$, with index $\i$ ($\j$) running along the $x$ ($y$) axis. $\tilde{\i}=\i+1$ and $\tilde{\j}=\j+1$. $\rho_{\i\j}$ is the conserved topological charge per plaquette $\i\j$, $j^x_{\i\j}$ ($j^y_{\i\j}$) is the flux per vertical (horizontal) link $\i\j$, which together satisfy the quantum continuity equation \rf{CE}.}
\label{sch}
\end{figure}

To construct a simple quantum theory, which reproduces the above classical hydrodynamics of vorticity in the limit of $\hbar\to0$, let us consider a square lattice model sketched in Fig.~\ref{sch}. We label each vertex of the lattice by two integer indices: $\i$ (along the $x$ axis) and $\j$ (along the $y$ axis). The same indices are used to label the square plaquettes, according to their lower left corner, as well as the vertical links going upward and the horizontal links to the right of the site $\i\j$. Each site contains a bosonic field $\Phi$ obeying the standard commutation algebra $[\Phi,\Phi^\dagger]=1$ (different sites commute).

We associate a charge density
\eq{
\rho_{\i\j}\equiv\frac{A^x_{\i\j}-A^x_{\i\tilde{\j}}+A^y_{\tilde{\i}\j}-A^y_{\i\j}}{2\pi a}
\label{rhoijA}}
to each plaquette, where $a$ is the lattice spacing. Here, $\tilde{\i}\equiv\i+1$ and $\tilde{\j}\equiv\j+1$, and
\eq{\al{
A^x_{\i\j}&=\frac{(\Phi^\dagger_{\tilde{\i}\j}+\Phi^\dagger_{\i\j})(\Phi_{\tilde{\i}\j}-\Phi_{\i\j})}{4ai}+{\rm H.c.}=\frac{\Phi^\dagger_{\i\j}\Phi_{\tilde{\i}\j}}{2ai}+{\rm H.c.}\,,\\
A^y_{\i\j}&=\frac{(\Phi^\dagger_{\i\tilde{\j}}+\Phi^\dagger_{\i\j})(\Phi_{\i\tilde{\j}}-\Phi_{\i\j})}{4ai}+{\rm H.c.}=\frac{\Phi^\dagger_{\i\j}\Phi_{\i\tilde{\j}}}{2ai}+{\rm H.c.}\,,
\label{QA}}}
which we assign formally to the corresponding horizontal and vertical sides of the plaquette, respectively. These definitions mimic Eqs.~\eqref{rho} and \eqref{A}, respectively, and should reproduce them by coarse graining the field configurations in the classical limit.

According to these conventions,
\eq{
\rho_{\i\j}=\frac{(\Phi^\dagger_{\tilde{\i}\j}-\Phi^\dagger_{\i\tilde{\j}})(\Phi_{\tilde{\i}\tilde{\j}}-\Phi_{\i\j})}{4\pi a^2i}+{\rm H.c.}\,.
\label{rhoij}}
We also see [from Eq.~\rf{rhoijA}] that
\eq{
Q=\sum_{\i\j}\rho_{\i\j}
}
vanishes in the bulk and reduces to the boundary terms, which we can interpret as the quantum version of the net vorticity \eqref{Q}. This suggests a conservation law with the boundary fluxes corresponding to the vorticity flow. Indeed, according to the Heisenberg equation of motion (for Hamiltonian $H$),
\eq{
\p_t\rho_{\i\j}=\frac{i}{\hbar}[H,\rho_{\i\j}]
}
can be seen to satisfy the continuity equation:
\eq{
\p_t\rho_{\i\j}+\frac{j^x_{\tilde{\i}\j}-j^x_{\i\j}+j^y_{\i\tilde{\j}}-j^y_{\i\j}}{a}=0\,.
\label{CE}}
Here, the fluxes are obtained by discretizing and quantizing the definition \rf{jmu}:
\eq{
j^x_{\i\j}=\frac{(\Phi^\dagger_{\i\tilde{\j}}-\Phi^\dagger_{\i\j})\p_t(\Phi_{\i\tilde{\j}}+\Phi_{\i\j})}{4\pi ai}+{\rm H.c.}\,,
\label{jij}}
and similarly for the other components. The time derivative should always be understood to denote the Heisenberg commutator:
\eq{
\p_t\mathcal{O}\equiv\frac{i}{\hbar}[H,\mathcal{O}]\,,
\label{EOM}}
for any (time-independent) operator $\mathcal{O}$.

It is useful to emphasize that this conservation law is not rooted in any specific symmetry of the system. Indeed, the form of the Hamiltonian $H$ still remains arbitrary. The continuity is rather dictated by the topology associated with the vorticity (hydro)dynamics in the interior of the system. Specifically, for a fixed field profile on the boundary, an arbitrary smooth field in the bulk yields the same net vorticity, irrespective of the details of the dynamics.

While the definitions \rf{QA} for the quantum field $\bA(\Phi)$ are motivated by the classical limit \rf{A}, which describes vorticity, any field $\bA(\Phi)$ entering Eq.~\rf{rhoijA} would in principle define a conserved dynamics. This is fully analogous to the arbitrary gauge field $\bA(\phi)$ parametrizing classical hydrodynamics associated with Eqs.~\rf{rho} and \rf{j}, as discussed above. The specific choice \rf{QA} is motivated by the classical correspondence to a physically meaningful extensive hydrodynamics in bosonic condensates.

\subsection{Boundary conditions}

The boundary conditions for a nonequilibrium injection of vorticity can be constructed based on energetics and general symmetry principles \cite{kimPRB15br,ochoaPRB16,kimPRL18}. In essence, the boundary-induced work can shift the energy barrier for a spontaneous injection of vorticity, in proportion to the applied bias. This bias can be established, for example, by a current applied in a metal contact tangentially to the interface \cite{grigorievaPRL04,kimPRB15br,ochoaPRB16} or a driven spin dynamics in an adjacent magnetic insulator \cite{kimPRL18}. Let us follow the latter scenario, supposing the magnetic order $\bn$ in the insulator couples to vorticity dynamics near the interface via a spin-orbit interaction. The relevant coarse-grained work $\delta W$ associated with a vorticity transfer $\delta Q$ across the interface then has an adiabatic contribution (at low frequencies) of the form \cite{kimPRL18}
\eq{
\delta W=g\,\bz\cdot\bn\times\dot{\bn}\,\delta Q\,.
\label{W}}
$g$ here is a phenomenological interfacial parameter for the coupling, $\bz$ is the normal to the ($xy$) plane of our bosonic film, and $\bn$ is taken to be the directional (unit-vector) order parameter of a ferromagnetic insulator. For $\bn$ steadily precessing around the $\bz$ axis, the effective bias becomes
\eq{
\mu\equiv\frac{\delta W}{\delta Q}=g\nu\Omega\,,
}
where $\Omega$ is the solid angle subtended by $\bn$.

This $\mu$ can be interpreted as establishing a local \textit{chemical potential for the vorticity,} supposing that the effective impedance for the vorticity transport is dominated by the bulk region. Physically, Eq.~\rf{W} describes the interfacial conversion of a pumped spin current (along the $z$ axis), $\propto\bz\cdot\bn\times\dot{\bn}$ \cite{tserkovRMP05} into the vorticity. We have explicitly derived the form of $g$, for a model of a ferromagnet/superconductor interface, in Ref.~\cite{kimPRL18}. The spin-to-vorticity interconversion described by Eq.~\rf{W}, however, can be expected to be general, as the $z$ component of spin and local vorticity transform similarly under the relevant structural (as well as time-reversal) symmetries.

\begin{figure}[!ht]
\includegraphics[width=\linewidth]{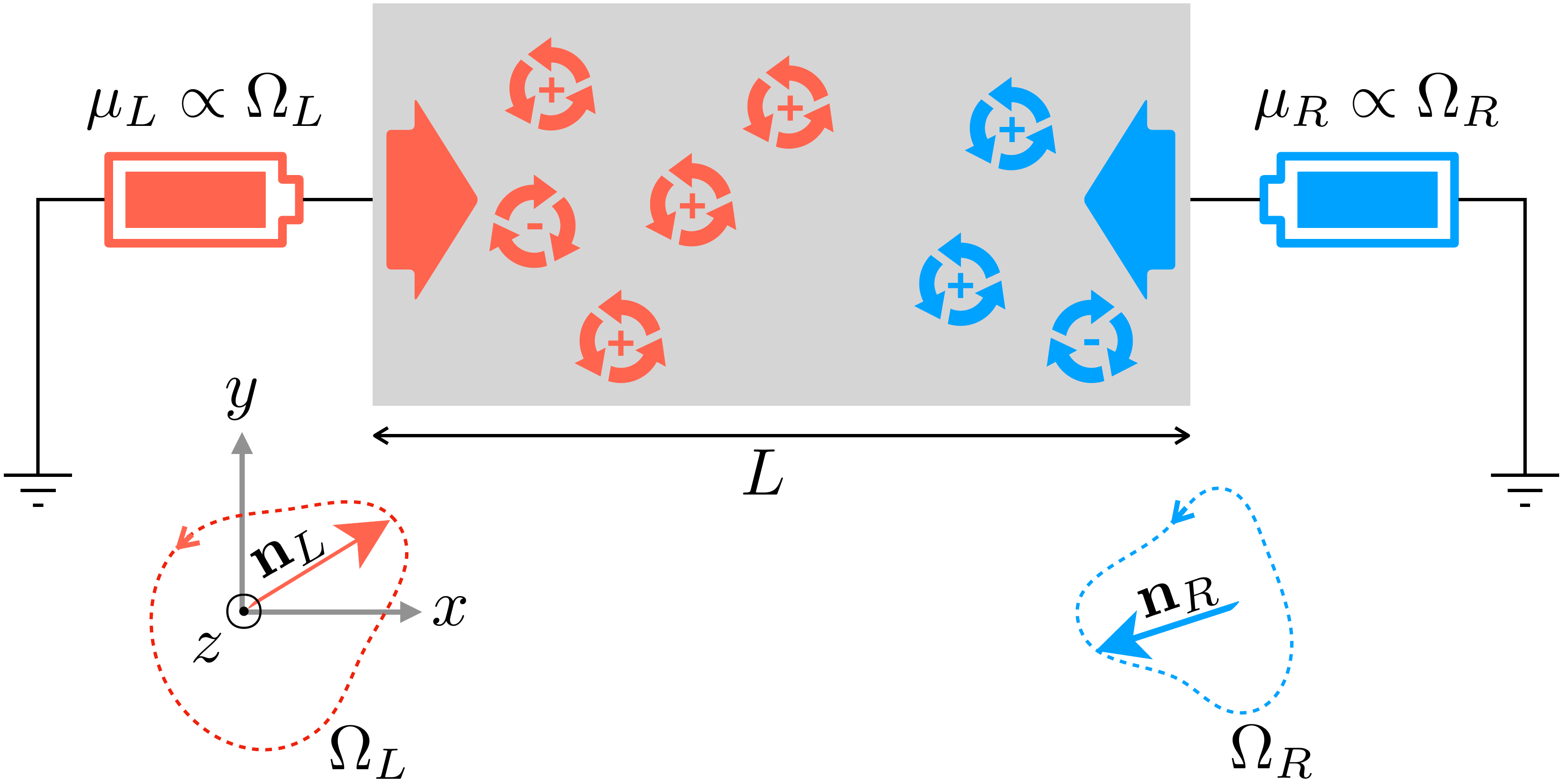}
\caption{Vorticity injection into a bosonic film. Coherently-precessing magnetic dynamics $\bn_L$ ($\bn_R$) at the left (right) side realizes a vorticity reservoir with a conjugate chemical potential $\mu_L$ ($\mu_R$). This effectively acts as a battery for the injection of the topological charge $Q$. A positive chemical potential leads to a build-up of a positive vorticity charge at the interface. If $\mu_L=\mu_R=\mu$ (which could be accomplished by attaching the same magnet symmetrically to both sides), an equilibrium state with the vortex chemical potential $\mu$ is established in the steady state, which has a vanishing vortex flux. If $\mu_L\neq\mu_R$, a dc vortex flux (driven by thermal and/or quantum fluctuations) is expected towards the lower chemical-potential side \cite{zouPRB19}.}
\label{vort}
\end{figure}

The natural chemical potential of vorticity in an equilibrium system (i.e., in the absence of magnetic dynamics $\dot{\bn}$) is $\mu\to0$, as the topological charge can freely go in and out of the vacuum. A circuit describing the out-of-equilibrium injection of vorticity into the bosonic medium by magnetic dynamics is sketched in Fig.~\ref{vort}. Particularly noteworthy is the case of $\mu_L=\mu_R=\mu$ in the figure, which corresponds to lifting the equilibrium chemical potential for the topological charge by the amount of $\mu$. This geometry is analogous to a rotating superfluid (where the precessing order parameter is replaced by the rotating container) and the frequency glitches in neutron-star pulsars (which have superfluid interiors surrounded by a rotating crust) \cite{haskellIJMPB15}.

\subsection{Kubo formula}

We are now ready to define the bulk impedance for the topological flow \rf{jmu}, as an intrinsic property of the bosonic system. Starting with a continuity equation for the coarse-grained quantum dynamics in the bulk, we have
\eq{
\p_t\rho+\BN\cdot\bj=0\,,
}
where the conserved density and current are obtained from Eqs.~\rf{rhoij} and \rf{jij}. We recall that the time derivatives are obtained in the Heisenberg picture. If we perturb the system by a scalar potential $\phi(\br,t)$ that couples to the topological charge, the Hamiltonian becomes
\eq{
H\to H+\int d^2r\,\phi(\br,t)\rho(\br)\,.
}
Note that the topological density \rf{rhoij} is even under time reversal, while the flux \rf{jij} is odd, so it vanishes in equilibrium, when $\phi\equiv0$. For a finite time-dependent potential $\phi$, on the other hand, the linear response is given by
\eq{
j_i(\br,t)=\int d^2r'dt'\chi_i(\br-\br',t-t')\phi(\br',t')\,,
}
where
\eq{
\chi_i(\br-\br',t-t')\equiv-i\theta(t-t')[j_i(\br,t),\rho(\br',t')]\,,
}
according to the Kubo formula (with the equilibrium expectation value implicit on the right-hand side).

To invoke the continuity equation, we differentiate the response function in time:
\begin{widetext}
\eq{\al{
\p_t\chi_i(\br-\br',t-t')&=i\theta(t-t')[j_i(\br,t),\p_{t'}\rho(\br',t')]-i\delta(t-t')[j_i(\br),\rho(\br')]\\
&=-i\theta(t-t')[j_i(\br,t),\BN'\cdot\bj(\br',t')]+\delta(t-t')\BN'\cdot\bp_i(\br-\br')\,,
}}
\end{widetext}
where the auxiliary curl-free function $\bp_i(\br-\br')$ is formally defined by inverting
\eq{
\BN'\cdot\bp_i(\br-\br')=-i[j_i(\br),\rho(\br')]\,.
\label{dp}}
We will see that it describes the response that is analogous to the paramagnetic component ($\propto i\varrho/m\omega$) of the electrical conductivity (for electrons of mass $m$ and density $\varrho$). Fourier transforming in time, $\bj(\omega)=\int dte^{i\omega t}\bj(t)$ etc., we finally get (summing over repeated indices)
\eq{
j_i(\br,\omega)=\frac{i}{\omega}\int d^2r'\,\chi_{ij}(\br-\br',\omega)\varepsilon_j(\br',\omega)\,.
}
Here,
\eq{\al{
\chi_{ij}(\br-\br',t-t')\equiv&-i\theta(t-t')[j_i(\br,t),j_j(\br',t')]\\
&+\delta(t-t')p_{ij}(\br-\br')
}}
is the current-current correlator ($p_{ij}\equiv \bp_i^{(j)}$) and
\eq{
\BE\equiv-\BN\phi
}
is the effective electric field. This gives for the conductivity tensor relating $\bj(\bk,\omega)$ to $\BE(\bk,\omega)$:
\eq{
\sigma_{ij}(\bk,\omega)=\frac{i}{\omega}\chi_{ij}(\bk,\omega)\,,
\label{jj}}
having also Fourier transformed in real space, $\int d^2r\,e^{-i\bk\cdot\br}$.

For the geometry sketched in Fig.~\ref{vort},
\eq{
\BE=g\nu\frac{\Omega_L-\Omega_R}{L}\bx\,,
}
supposing that the length of the topological transport channel $L$ is long enough, so that the bulk dominates over the interfacial impedances \cite{tserkovJAP18}. We take $g$ and $\nu$ to be the same at the two interfaces. Note that the conductivity should generally depend on the topological chemical potential $\mu$, which can be controlled by the average dynamic bias, $\Omega_L+\Omega_R$. We thus conclude that the sum $\Omega_L+\Omega_R$ effectively gates the bosonic vorticity conduit, while the difference $\Omega_L-\Omega_R$ establishes a topological flux through it. As the conductivity tensor $\hat{\sigma}$ can be exponentially sensitive to $\mu$ at low temperatures, this suggests a potential transistor functionality.

\subsection{Electrical transconductance}

Having established the vorticity response to the magnetic dynamics in the structure like that sketched in Fig.~\ref{vort}, we can now consider its nonlocal feedback on the magnetic dynamics. To that end, we invoke the Onsager reciprocity, in order to establish the torque induced by the vorticity flow through the interfaces \cite{kimPRL18}:
\eq{
\BT=g\,\bn\times\bz\times\bn\,j=\theta\,\bn\times\bz\times\bn\,,
}
where $j$ is the vorticity flux impinging on the magnetic insulator. This is known as the (anti)damping-like torque, which plays an important role in spin-torque-induced magnetic dynamics \cite{ralphJMMM08}. In particular, at a critical value of its magnitude $\theta$, the ferromagnet can undergo an instability driving it into coherent self-oscillations. The magnitude of such a torque acting on the right magnet due to a coherent resonant dynamics (at frequency $\nu$) induced in the left magnet is given by
\eq{
\theta=gj=\frac{g^2\nu\Omega}{L}\sigma_{xx}
}
We recall that $g$ is a phenomenological parameter of the interface, whose existence is dictated by structural symmetries and whose magnitude depends on the details of the interfacial coupling (including corrections due to quantum fluctuations). The longitudinal conductivity $\sigma_{xx}$ reflects the intrinsic vorticity transport across the bosonic lattice.

In the particle-superfluid limit, when the vorticity is carried by the plasma of solitonic defects with quantized topological charge $\pm1$ and mobility $M$, the corresponding conductivity is simply $\sigma_{xx}=2\rho M$, where $\rho$ is the density of the unbound vortex-antivortex pairs (well above the Kosterlitz-Thouless transition) \cite{zouPRB19}. The associated diffusion coefficient is given by $D=k_BTM$, according to the Einstein-Smoluchowski relation. For large vortices, the mobility may be limited by the dissipation associated with the normal-fluid component (which is perturbed by the vortex motion). Reference~\cite{kimPRL18} offers some quantitative estimates of the torques induced by vortex motion in high-temperature superconducting films, suggesting its practical relevance.

\subsection{Superfluidity of vorticity}

Exploiting the particle-vortex duality \cite{wenBOOK04}, we consider a situation when a strong interparticle repulsion prevents the ordinary mass flow. In this case, an insulating state for the particle dynamics may exhibit superfluidity for the topological charge (i.e., vorticity). To this end, we pursue an effective description with the Hamiltonian density
\eq{
H=\frac{\rho^2}{2\chi}+\frac{\mA(\BN\psi)^2}{2}\,,
}
expressed in terms of the coarse-grained (condensed) vorticity density $\rho$ and its condensate phase $\psi$. $\chi$ is the thermodynamic compressibility of vorticity and $\mA$ is the phase stiffness. This form of the Hamiltonian, along with the conjugacy relation $[\psi(\br),\rho(\br')]=i\delta(\br-\br')$, reflects an emergent gauge structure associated with the global conservation of vorticity.

The associated flux can be read out from the Hamilton equation for the density dynamics:
\eq{
\hbar\p_t\rho=-\p_\psi H=\mA\nabla^2\psi~~~\Rightarrow~~~\bj=-\mA\BN\psi/\hbar\,.
}
Phase dynamics is described by the Josephson relation:
\eq{
\hbar\p_t\psi=\p_\rho H=\rho/\chi\,.
}
The mean-field current self-correlator can be found as the current response to perturbation $H\to H+\bs\cdot\bj$, which modifies the equation of motion as:
\eq{
\hbar\p_t\rho=\mA\nabla^2\psi-\mA\BN\cdot\bs/\hbar\,.
}
The long-wavelength response thus vanishes, as $\BN\cdot\bs=0$.

We are therefore left with evaluating the ``paramagnetic" contribution, which follows from Eq.~\rf{dp}. The associated current-density correlator
\eq{
[\bj(\br),\rho(\br')]=-i\mA\BN\delta(\br-\br')/\hbar
}
gives
\eq{
p_{ij}=\mA\delta(\br-\br')\delta_{ij}/\hbar\,,
}
resulting in the diagonal dynamic (long-wavelength) conductivity
\eq{
\sigma(\omega)=\frac{i\mA}{\hbar\omega}\,.
}
Regularizing this result at zero frequency, $\omega\to\omega+i0^+$, we get ${\rm Re}\sigma=(\pi\mA/\hbar)\delta(\omega)$. As expected, the static conductivity diverges in the low-frequency limit. In this case, the superfluid bulk has no impedance and the vorticity conductance of the entire structure needs to be determined by carefully considering the interfacial injection physics, which is akin to the Andreev conductance of normal/superconducting interfaces \cite{nazarovBOOK09}.

\section{Summary and outlook}
\label{so}

Motivated by the conceptual attraction of solid-state transport phenomena emerging out of real-space topological invariants \cite{tserkovJAP18}, we set out to construct a field-theoretic Kubo formalism for evaluating the associated transport coefficients. Two basic issues arise in this regard: (1) The underlying topological invariants typically appear at the level of a coarse-grained classical description that is, furthermore, projected onto a low-energy manifold, in the spirit of the Landau order-parameter formulation; and (2) related to this, there are generally dynamical processes that allow for rapid transitions (``phase slips") between different topological sectors of the theory, which may be driven by classical and/or quantum fluctuations.

In this paper, we showed that a topological conservation law may also arise at the most microscopic quantum level, without a need for any higher-level Landau-type coarse graining. The conservation law here is distinct from the more conventional examples of the topological hydrodynamics \cite{tserkovJAP18}, due to the existence of the bulk-edge correspondence (such as the bulk vorticity vs edge winding) rooted in a variant of a Stokes theorem. The nonlocal topological character of the ensuing extensive bulk hydrodynamics engenders a robust continuity equation that is immune to any local fluctuations. Arbitrary global (thermal and quantum) fluctuations, furthermore, are fully accounted for by the topological charge fluxes across the boundaries, which, in turn, offer means for injecting and detecting the bulk hydrodynamics.

A general approach for constructing a practical device, in which the transport coefficients associated with this topological hydrodynamics may be measured, can be implemented based on the energetics and thermodynamic reciprocities of the nonequilibrium response. This allows us to formulate a Kubo linear-response approach both for calculating and for measuring the topological charge conductivity. As has been recently illustrated by measuring the electrical transconductance induced by winding dynamics of a hidden magnetic N{\'e}el order \cite{takeiPRL16,*stepanovNATP18}, these ideas may broaden the scope of transport-based investigations of fundamental correlations and ordering in quantum materials. We propose, in particular, to utilize a topological transport probe to test the purported particle-vortex duality of the vortex-superfluid (i.e., particle-insulator) side of the superfluid-insulator quantum phase transition.

\begin{acknowledgments}
We are grateful to Se Kwon Kim for insightful discussions. The work was supported by the U.S. Department of Energy, Office of Basic Energy Sciences under Award No.~DE-SC0012190.
\end{acknowledgments}


%

\end{document}